\documentclass[conference,final]{IEEEtran}

\ifCLASSINFOpdf
\else
\fi

\usepackage[cmex10]{amsmath}
\usepackage{amsfonts}
\usepackage{amssymb}
\usepackage{graphicx}

\begin{document}

\author{
\IEEEauthorblockN{Gerhard Kramer and Jie Hou}
\IEEEauthorblockA{%Fakult\"at f\"ur Elektrotechnik und Informationstechnik \\
Technische Universit\"at M\"unchen\\
Arcisstra{\ss}e 21, 80333 M\"unchen, Germany\\
gerhard.kramer@tum.de, jie.hou@tum.de}
%\and
%\IEEEauthorblockN{Jie Hou}
%\IEEEauthorblockA{Fakult\"at f\"ur Elektrotechnik und Informationstechnik \\
%Technische Universit\"at M\"unchen\\
%Arcisstra{\ss}e 21, 80333 M\"unchen, Germany\\
%jie.hou@tum.de}
%\thanks{abc}
}

\IEEEoverridecommandlockouts
\IEEEpeerreviewmaketitle

%=============================================================
\title{Short-Message Quantize-Forward Network Coding }

\maketitle

\begin{abstract}
Recent work for single-relay channels shows that quantize-forward (QF) with long-message encoding achieves the same reliable rates as compress-forward (CF) with short-message encoding. It is shown that short-message QF with backward or pipelined (sliding-window) decoding also achieves the same rates. Similarly, for many relays and sources, 
short-message QF with backward decoding achieves the same rates as long-message QF.  Several practical advantages of short-message encoding are pointed out, e.g.,  reduced delay and simpler modulation. Furthermore, short-message encoding lets relays use decode-forward (DF) if their channel quality is good, thereby enabling multi-input, multi-output (MIMO) gains that are not possible with long-message encoding. Finally, one may combine the advantages of long- and short-message encoding by hashing a long message to short messages.
\end{abstract}

%=============================================================
%=============================================================
\section{Introduction}
\label{sec:intro}
Relaying is receiving attention for wireless cellular applications because it improves rates and reliabilities. There are two simple geometric scenarios that give insight into relaying strategies, and that show how relaying achieves distributed multi-input, multi-output (MIMO) gains~\cite{Gastpar02}. First, relays that are close to a source node can achieve ``multi-input" gains by using a decode-forward (DF) strategy. Second, relays that are close to a destination node can achieve ``multi-output" gains by using a compress-forward (CF) strategy. Both the DF and CF strategies appeared for abstract channels in the work of Cover and El Gamal~\cite{Cover79}. This document focuses on the CF strategy whose usefulness for network communication has been demonstrated., e.g., in~\cite{Schein00,Ozgur07} and follow-up works.

Recently, a method called {\it noisy network coding} was developed~\cite{Avestimehr09,Lim10} that is a quantize-forward (QF) variant of the CF strategy. The QF strategy uses simple relays and achieves a remarkably simple-to-describe rate region that is sometimes close to a cut-set upper bound. Our main goal is to outline why the short-message CF method of~\cite[Sec. V]{KGG05} can achieve better rates and reliabilities with less complexity than the long-message QF methods of~\cite{Avestimehr09,Lim10}. This extends results by Wu and Xie~\cite{WuXie10,WuXie11} who found that short-message encoding achieves the same rates as long-message encoding for a single source. Finally, we describe a hashing scheme that combines some of the advantages of long- and short-message encoding.

%=============================================================
\section{Taxonomy} \label{sec:taxonomy}

We first address terminology. Variations of the compression strategy of Cover and El Gamal~\cite{Cover79} are known by names such as ``estimate-forward" (EF), ``compress-forward" (CF), ``quantize-forward" (QF), ``quantize-map", ``hash-forward" (HF), and so forth. We make the following observations.
\begin{itemize}
\item The word ``compress" is a generic name that refers to both lossless and lossy source coding, the latter including ``quantization" and ``hashing" (or ``binning").
\item Without hashing one obtains a QF strategy~\cite{Ozgur07,Avestimehr09,Lim10}.
\item Without quantization one obtains a HF strategy~\cite{Kim08}.
\item The name of a relay function should not depend on the choice of operations at other nodes. In particular, it should not depend on whether other nodes perform optimal or suboptimal processing.
\end{itemize}

The last bullet point sometimes causes confusion. Some literature takes CF to mean that the ``next" relay along a route {\it must decode} certain indices, perhaps even with a suboptimal decoder. However, if one accepts this bullet point, such terminology makes little sense. We therefore advocate to use the (generic) name CF for the general strategy, HF for a strategy without quantization, and QF for a strategy without hashing. Of course, this makes HF and QF (and noisy network coding) special cases of CF. 

%===================================================================================================================================
\section{Quantization Suffices} \label{sec:QF}
We review the recent QF strategy, see~\cite{Avestimehr09,Lim10}. However, rather than using long-message repetition encoding we use ``short"-message encoding (see~\cite[Thm.~6]{Cover79},~\cite[Sec. V]{KGG05},~\cite{WuXie10,WuXie11,Kramer09}) and pipelined decoding via a sliding-window method (see~\cite[Sec.~I.A]{KGG05},~\cite{Kramer09},~\cite[p.~842]{Carleial82},~\cite[p.~761]{Xie04},).
 As usual, we use independent random codebooks for each block (see~\cite[p.~842]{Carleial82} and~\cite[p.~760]{Xie04}).  We use the notation $x^n=x_1,x_2,\ldots,x_n$ and $T_{\epsilon}^n(P_X)$ for $\epsilon$-typical sets.

\begin{figure}[t!]
  \centerline{\includegraphics[scale=0.5]{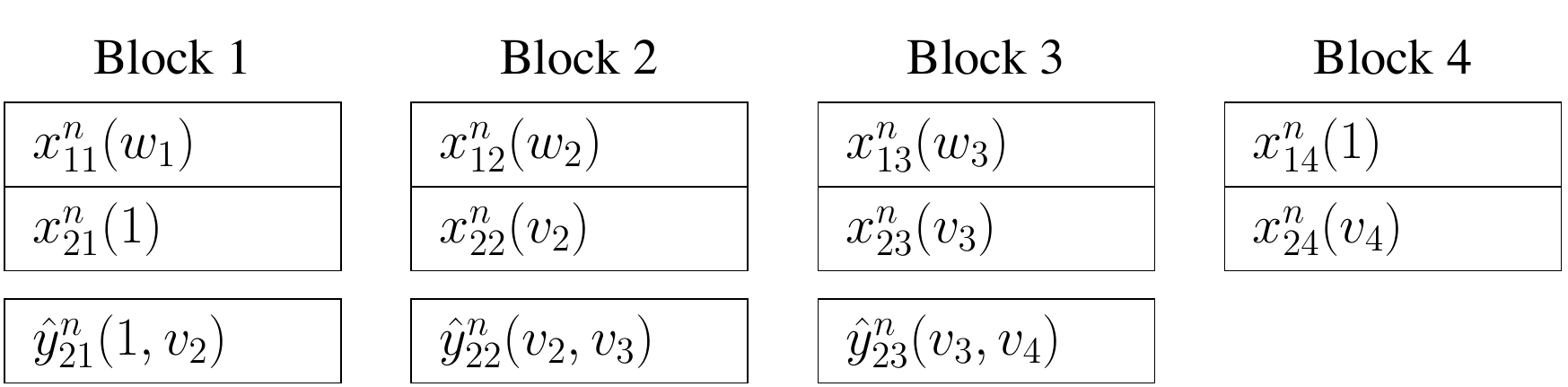}}
  \caption{A quantize-forward strategy for the relay channel.}
  \label{fig:relay1-compressForward}
\end{figure}

\medskip\noindent {\textit{Code Construction}}: Encoding is performed in $B+1$ blocks,
and we generate a different code book for each block
(see \hbox{Figure}~\ref{fig:relay1-compressForward} where $B+1=4$).
For block $b$, $b=1,2,\ldots,B+1$, generate $2^{nR}$ codewords $x_{1b}^n(w)$,
$w=1,2,\ldots,2^{nR}$, by choosing the symbols $x_{1bi}(w)$, $i=1,2,\ldots,n$, independently
using $P_{X_1}(\cdot)$. Similarly, generate $2^{nR_2}$ codewords
$x_{2b}^n(v)$, $v=1,2,\ldots,2^{nR_2}$, by choosing the $x_{2bi}(v)$
independently using $P_{X_2}(\cdot)$. Finally, introduce an auxiliary
random variable $\hat{Y}_2$ that represents a quantized
version of $Y_2$, and consider a distribution
$P_{\hat{Y}_2|X_2}(\cdot)$. For each $x_{2b}^n(v)$, generate
$2^{nR_2}$ codewords $\hat{y}_{2b}^n(v,u)$, $u=1,2,\ldots,2^{nR_2}$, by
choosing the $\hat{y}_{2bi}(v,u)$ independently using
$P_{\hat{Y}_2|X_2}(\cdot|x_{2bi}(v))$.

%%%%%%%%%%%%%%%%%%%%%%%%%%%%%%%%%%%%%%%%
%%%%%%%%%%%%%%%%%%%%%%%%%%%%%%%%%%%%%%%%

\medskip\noindent {\textit{Source}}: The message $w$ of $2^{nRB}$ bits is split into
$B$ blocks $w_1,w_2,\ldots,w_B$ of $2^{nR}$ bits
each. In block $b$, $b=1,2,\ldots,B+1$, the source transmits
$x_{1b}(w_b)$, where $w_{B+1}=1$.

\medskip\noindent {\textit{Relay}}: In block~$b=1$, the relay transmits
$x_2^n(1)$. After block $b$, the relay has seen $y_{2b}^n$.  The relay
tries to find a $\tilde{u}_b$ such that
\begin{align}
  (\hat{y}_{2b}^n(v_b,\tilde{u}_b), x_{2b}^n(v_b),y_{2b}^n)
  \in T_\epsilon^n(P_{\hat{Y}_2X_2Y_2}).
  \label{eq:relay1-CFrelayDecoder}
\end{align}
If one or more such $\tilde{u}_b$ are found, then the
relay chooses one of them, sets $v_{b+1}=\tilde{u}_b$, and transmits
$x_{2(b+1)}(v_{b+1})$. If no such pair is found, the relay sets
$v_{b+1}=1$ and transmits $x_{2(b+1)}(1)$.

\medskip\noindent {\textit{Sink Terminal}}: After block $b$, $b=2,3,\ldots,B+1$, the receiver
has seen $y_{3(b-1)}^n$ and $y_{3b}^n$, and
tries to find a pair $(\tilde{w}_{b-1},\tilde{v}_b)$ such that
\begin{align}
    & (x_{2b}^n(\tilde{v}_b),y_{3b}^n) \in T_\epsilon^n(P_{X_2Y_3}) \text{ and} \label{eq:sinkDecoder1} \\
    & (x_{1(b-1)}^n(\tilde{w}_{b-1}),\hat{y}_{2(b-1)}^n(\hat{v}_{b-1},\tilde{v}_{b}),
     x_{2(b-1)}^n(\hat{v}_{b-1}), y_{3(b-1)}^n) \nonumber \\
    & \in T_\epsilon^n(P_{X_1\hat{Y}_2X_2Y_3}),
  \label{eq:sinkDecoder2}
\end{align}
and we assume that $\hat{v}_{b-1}=v_{b-1}$.
If one or more such $(\tilde{w}_{b-1},\tilde{v}_b)$ are found, then the sink chooses one
of them, and puts out this choice as $(\hat{w}_{b-1},\hat{v}_{b})$.  If no such
$(\tilde{w}_{b-1},\tilde{v}_b)$ is found, the sink puts out $(\hat{w}_{b-1},\hat{v}_{b})=(1,1)$.

\medskip\noindent{\textit{Analysis}}: The analysis follows familiar steps, see~\cite[Sec. 15.2]{Cover06}
and we summarize the results.
\begin{enumerate}
   \item The relay quantization requires
\begin{align}
   R_2> I(\hat{Y}_2 ; Y_2 | X_2).
   \label{eq:relayBound}
\end{align}

   \item The sink's decoder can be viewed as a multi-access channel (MAC) decoder for
   two messages $w_{b-1}$ and $v_b$ and therefore we have the bounds
\begin{align}
   & R < I(X_1 ; \hat{Y}_2 Y_3 | X_2)  \label{eq:sinkBound1} \\
   & R_2 < I(X_2;Y_3) + I(\hat{Y}_2 ; X_1 Y_3 | X_2) \label{eq:sinkBound2} \\
   & R+R_2 %< I(X_2;Y_3) + I(X_1;\hat{Y}_2 ; Y_3 | X_2) \label{eq:sinkBound3} \\
   < I(X_1 X_2;Y_3) + I(\hat{Y}_2 ; X_1 Y_3 | X_2). \label{eq:sinkBound3}
\end{align}
%  where
%\begin{align}
%  & I(A;B;C|D)= \nonumber \\
%  & H(A|D)+H(B|D)+H(C|D) - H(A B C | D).
%\end{align}
Observe that we {\it cannot} ignore the bound \eqref{eq:sinkBound2}, as might be expected, because we
require that $\hat{v}_{b-1}=v_{b-1}$ in \eqref{eq:sinkDecoder2}. The sums in
\eqref{eq:sinkBound2} and \eqref{eq:sinkBound3} are due to the intersection of independent events 
\eqref{eq:sinkDecoder1} and \eqref{eq:sinkDecoder2}.
\end{enumerate}

The joint distribution of the random variables factors as
\begin{align}
   P_{X_1}(a) P_{X_2}(b) P_{Y_2Y_3|X_1X_2}(c,d|a,b) P_{\hat{Y}_2|X_2Y_2}(f|b,c)
   \label{eq:relay1-CFPDF}
\end{align}
for all $a,b,c,d,f$. Performing a Fourier-Motzkin elimination of $R_2$, and manipulating
the mutual information expressions, the bounds \eqref{eq:relayBound}-\eqref{eq:sinkBound3}
become
\begin{align}
   & R < I(X_1 ; \hat{Y}_2 Y_3 | X_2)  \label{eq:sinkBound1a} \\
   & R < I(X_1 X_2 ; Y_3) -  I(\hat{Y}_2 ; Y_2 | X_1 X_2 Y_3) \label{eq:sinkBound2a} \\
   & I(\hat{Y}_2 ; Y_2 | X_1 X_2 Y_3) < I(X_2 ; Y_3) \label{eq:sinkBound3a}
\end{align}
But suppose that \eqref{eq:sinkBound3a} is not satisfied so that
\eqref{eq:sinkBound2a} and \eqref{eq:sinkBound3a} give
\begin{align}
   & R < I(X_1 ; Y_3 | X_2)  \label{eq:sinkBound1b}
\end{align}
which is a stronger bound than \eqref{eq:sinkBound1a}.
The rates satisfying \eqref{eq:sinkBound1b} are achievable with QF, e.g., by choosing $\hat{Y}_2$ independent
of $X_2$ and $Y_2$ (and thus $X_1$ also). Hence we may ignore the constraint \eqref{eq:sinkBound3a}.
The resulting QF rates are as close as desired to the known CF rate
\begin{align}
   & R_{CF} = \max \min \left[ I(X_1 ; \hat{Y}_2 Y_3 | X_2), \right.  \nonumber \\
   & \qquad \qquad \left. I(X_1 X_2 ; Y_3) -  I(\hat{Y}_2 ; Y_2 | X_1 X_2 Y_3) \right] \label{eq:CFrate}
\end{align}
where the maximization is over all distributions factoring as in \eqref{eq:relay1-CFPDF}. The rate \eqref{eq:CFrate} is the
same as the more commonly used expression (see~\cite[Thm. 3 and eq. (6)]{ElGamal10})
\begin{align}
   & R_{CF} = \max I(X_1 ; \hat{Y}_2 Y_3 | X_2) \nonumber \\
   & \text{subject to }  I(\hat{Y}_2 ; Y_2 | X_2 Y_3) \le I(X_2 ; Y_3). \ \label{eq:CFrate1}
\end{align}

%===================================================================================================================================
\section{Discussion} \label{sec:discussion}

%We discuss extensions of the above methods, and we compare long- and short-message encoding.

%-------
\subsection*{Short Messages and Backward Decoding}
One may decode short-message encoded packets by using backward decoding (see, e.g.,~\cite{KGG05,WuXie11,WillemsThesis}) as long as the final transmission block is sufficiently long to be able to decode $v_{B+1}$. The bound \eqref{eq:sinkBound2} is replaced with the weaker constraint
\begin{align}
   & R_2 < I(X_2;Y_3 | X_1) + I(\hat{Y}_2 ; X_1 Y_3 | X_2) \label{eq:sinkBound2d}
\end{align}
which means that \eqref{eq:sinkBound3a} is replaced with the weaker constraint
\begin{align}
   & I(\hat{Y}_2 ; Y_2 | X_1 X_2 Y_3) < I(X_2 ; Y_3 | X_1) \label{eq:sinkBound3d}.
\end{align}
We thus find that backward decoding should outperform pipelined decoding for slow-fading channels with outage.

Another possibility is for the receiver to jointly decode all indices after transmission is completed.
Yet another possibility is to use a pipelined decoder with a longer and variable window length, either in the forward or backward directions.
For example, the window length may be $b$ for block $b$.

%-------
\subsection*{Advantage of Long Messages: Reliability}
One advantage of long-message encoding is that the quantization constraint \eqref{eq:sinkBound2} disappears (which means that \eqref{eq:sinkBound3a} disappears). Thus, long-message encoding should outperform short-message encoding for slow-fading channels with outage.

%------
\subsection*{Advantage of Short Messages: DF and Distributed MIMO}
As mentioned in the introduction, relaying achieves distributed MIMO gains if relays close to a source use DF and relays far from a source use CF/QF~\cite{Gastpar02,Ozgur07}. Unfortunately, long-message encoding inhibits DF because the message is usually too long to decode after receiving one block of channel outputs. In contrast, short-message encoding lets relays close to a source decode messages early. These relays can form a distributed transmit array with this source.

%------
\subsection*{Advantage of Short Messages: Modulation Complexity}
A subtle advantage of short messages is that one can map a small number of bits onto the modulation, i.e., the modulation set can be kept small. When using long-message encoding, one would either have to use a (very) large modulation set, or one must first hash the long message to a shorter message (see below). Either way, long-message decoding is complex and will suffer implementation and synchronization losses if a large modulation alphabet is used.

%-------
\subsection*{Advantage of Short Messages: Encoding and Decoding Delay}
Short-message encoding has a considerably-reduced {\it encoding} delay as compared to long-message encoding. Similarly, pipelined decoding enjoys a considerably-reduced {\it decoding} delay. The combination of these two approaches might support streaming applications.

%------
\subsection*{Alternative: Hashing a Long Message to Short Messages}
We may combine the reliability advantage of long-message encoding with the early decoding and modulation complexity advantages of short messages by hashing the long message to many short messages before transmission. Relays may now decode short hash messages if they can, thereby enabling DF and distributed MIMO. Furthermore, the quantization constraints such as \eqref{eq:sinkBound2} disappear like for long-message encoding. The price paid is long encoding and decoding delays.

%------
\subsection*{Multiple Relays, Messages, and Destinations}
As shown in~\cite{WuXie10,WuXie11}, short-message encoding and backward decoding achieves the same bounds on $R$ as in~\cite{Avestimehr09,Lim10} for multiple relays. This statement is also valid for multiple relays and multiple sources. %It turns out that one must additionally consider many constraints of the form \eqref{eq:sinkBound2}, but the final result (the achievable rates) does not change.
%Pipelined decoding does not seem to achieve the same rates as long-message encoding and decoding (or short-message encoding and backward decoding).

%-------
\subsection*{Quantizing and Hashing}
The knowledgable reader may wonder why hashing (or binning) is not needed to achieve the CF rates, in seeming contradiction to results in, e.g.,~\cite{Kim08} and~\cite
{KimSkoglundCaire09}. Of course, one obvious explanation is that we are using a better (joint) decoder rather than a step-by-step decoder. 

However, the model of~\cite{Kim08} deserves closer inspection. The relay channel in~\cite{Kim08} does not have the ``standard" form with a memoryless channel $p(y_2,y_3|x_1,x_2)$; there is instead a rate constraint $R_0$ on the relay-destination link. But we can bring such a channel into standard form by introducing a random variable $X_2$ that represents the relay's transmit symbols and choose its alphabet size $|\mathcal{X}_2|$ as $2^{R_0}$ (if $2^{R_0}$ is not an integer we may choose a model with memory on the relay-destinaton link and again appropriately limit the size of the input alphabet). Now suppose that $R_2>R_0$ in which case QF necessarily assigns the same codeword $x_{2b}^n$ to (exponentially in $n$) many indices $v_b$. In other words, QF implicitly performs hashing. The same consideration shows that Wyner-Ziv coding may be considered to be using QF only (without a binning step). A similar claim can be made for Slepian-Wolf coding. Of course, this statement lacks depth since whether we call implicit binning QF or HF is not important.

The reader may now wonder whether QF always performs hashing implicitly. We emphasize that this is generally {\it not} the case when $R_2<\log |\mathcal{X}_2|$. For example, for real-input channels such as Gaussian channels we have $|\mathcal{X}_2|=\infty$ and QF will generally assign a unique $x_{2b}^n$ to every index $v_b$.

%===================================================================================================================================
\section{Conclusion} \label{sec:conclusion}
For the single-relay channel, short-message QF with pipelined decoding achieves the same rates as long-message QF. For the multi-relay, multi-source channel, short-message QF with backward decoding recovers the rates of long-message QF. Several advantages of short-message encoding are pointed out, e.g., substantial reduction in delay, reduced modulation complexity, and added flexibility in letting relays choose DF or CF (or QF).

%%%%%%%%%%%%%%%%%%%%%%%%%%%%%%%%%%%%%%%
\section*{Acknowledgment}
\label{sec:ack}
%%%%%%%%%%%%%%%%%%%%%%%%%%%%%%%%%%%%%%%%
G. Kramer was supported by an Alexander von Humboldt Professorship endowed by the German Federal Ministry of Education and Research and by NSF Grant CCF-09-05235. 
Jie Hou was supported by NEWCOM++.

%===================================================================================================================================

%\appendix
%===================================================================================================================================
%\section*{Proof}
%
%Appendix

%\bibliographystyle{plain}
%\bibliographystyle{IEEE}
%\bibliographystyle{unsrt}
%\bibliography{ITgeneral,mu}

\end{document}